%% file: main.tex
\documentclass[sigconf,10pt,balance=false]{acmart}

\usepackage{tabularx}
\renewcommand\footnotetextcopyrightpermission[1]{}

\input{my_def}

\usepackage{enumitem}

\newcommand{\ie}{\emph{i.e.,}\xspace}

\begin{document}

\title{\framework{}: A Robust System for Spoofing User Data in Android Platforms}

\author{
  Harish Yadav \quad
  Vikas Maurya \quad
  Abhilash Jindal \quad
  Vireshwar Kumar
}

\email{
csy217544@iitd.ac.in}  \email{vikasmauryaphd@gmail.com} \email{ajindal@cse.iitd.ac.in} \email{viresh@cse.iitd.ac.in}

\affiliation{%
  \institution{Indian Institute of Technology Delhi}
}

%


\input{0_abstract}
\settopmatter{printfolios=true}
\maketitle

\input{1_introduction}
\input{8_related_work}

\input{2_motivation}
\input{3_methodology}
\input{4_architecture}
\input{6_mitigating_side_channel_attacks}
\input{5_protecting_user_privacy}
\input{7_results}
\input{9_conclusion}

\bibliographystyle{unsrt}
\bibliography{10_references}



\end{document}

%% file: my_def.tex
\usepackage{xspace}
\usepackage{algorithmic}
\usepackage{graphicx}
\usepackage{textcomp}
\usepackage{subcaption}
\usepackage{array}
\usepackage{makecell}
\usepackage{listings}
\usepackage{multirow}

\usepackage{tikz}
\newcommand*\circled[1]{\tikz[baseline=(char.base)]{
            \node[shape=circle,draw,inner sep=1pt] (char) {#1};}}
            
\usepackage{hyperref}

\input{listing_style}
 
\newcolumntype{M}[1]{>{\centering\arraybackslash}m{#1}}

\newcolumntype{L}[1]{%
 >{\vbox to 2ex\bgroup\vfil\arraybackslash}%
 m{#1}%
 <{\egroup}}

\newcolumntype{P}[1]{>{\centering\arraybackslash}p{#1}}

\newcommand{\framework}{WhiteLie\xspace}
\newcommand{\highlightedPositive}[1]{\color{teal}\textbf{#1}\color{black}}
\newcommand{\highlightedNegative}[1]{\color{red}#1\color{black}}

\usepackage{pifont}
\newcommand{\cmark}{\highlightedPositive{\ding{51}}}%
\newcommand{\xmark}{\highlightedNegative{\ding{55}}}%

%% file: listing_style.tex
\definecolor{codegreen}{rgb}{0,0.6,0}
\definecolor{codegray}{rgb}{0.5,0.5,0.5}
\definecolor{codepurple}{rgb}{0.58,0,0.82}
\definecolor{backcolour}{rgb}{0.96,0.96,0.96}

\lstdefinestyle{mystyle}{
    backgroundcolor=\color{backcolour},   
    commentstyle=\color{codegreen},
    keywordstyle=\color{magenta},
    numberstyle=\tiny\color{codegray},
    stringstyle=\color{codepurple},
    basicstyle=\ttfamily\footnotesize,
    breakatwhitespace=false,         
    breaklines=true,                 
    captionpos=b,                    
    keepspaces=true,                 
    numbers=left,                    
    numbersep=5pt,                  
    showspaces=false,                
    showstringspaces=false,
    showtabs=false,                  
    tabsize=2
}

\lstdefinelanguage{Kotlin}{
  comment=[l]{//},
  commentstyle={\color{gray}\ttfamily},
  emph={filter, first, firstOrNull, forEach, lazy, map, mapNotNull, println},
  emphstyle={\color{red}},
  identifierstyle=\color{black},
  keywords={!in, !is, abstract, actual, annotation, as, as?, break, by, catch, class, companion, const, constructor, continue, crossinline, data, delegate, do, dynamic, else, enum, expect, external, false, field, file, final, finally, for, fun, get, if, import, in, infix, init, inline, inner, interface, internal, is, lateinit, noinline, null, object, open, operator, out, override, package, private, property, protected, public, receiveris, reified, return, return@, sealed, set, setparam, super, suspend, tailrec, this, throw, true, try, typealias, typeof, val, var, vararg, when, where, while},
  keywordstyle={\color{blue}\bfseries},
  morecomment=[s]{/*}{*/},
  morestring=[b]",
  morestring=[s]{"""*}{*"""},
  ndkeywords={@Deprecated, @JvmField, @JvmName, @JvmOverloads, @JvmStatic, @JvmSynthetic, @Throws, Array, Byte, Double, Float, Int, Integer, Iterable, Long, Runnable, Short, String, Any, Unit, Nothing},
  ndkeywordstyle={\color{orange}\bfseries},
  sensitive=true,
  stringstyle={\color{teal}\ttfamily},
}

%% file: 0_abstract.tex
\begin{abstract}
Android employs a permission framework that empowers users to either accept or deny sharing their private data (e.g., location) with an app. However, many apps tend to crash when they are denied permission, leaving users no choice but to allow access to their data in order to use the app. In this paper, we introduce a comprehensive and robust user data spoofing system, \framework{}, that can spoof a variety of user data and feed it to target apps. Additionally, it detects privacy-violating behaviours, automatically responding by supplying spoofed data instead of the user’s real data, without crashing or disrupting the apps. Unlike prior approaches, \framework{} requires neither device rooting nor altering the app's binary, making it deployable on stock Android devices. Through experiments on more than 70 popular Android apps, we demonstrate that \framework{} is able to deceive apps into accepting spoofed data without getting detected. Our evaluation further demonstrates that \framework{} introduces negligible overhead in terms of battery usage, CPU consumption, and app execution latency. Our findings underscore the feasibility of implementing user-centric privacy-enhancing mechanisms within the existing Android ecosystem.
\end{abstract}

\keywords{Android Security, Mobile Privacy, Data Spoofing, Permission Abuse, Side-Channel Attacks}

%% file: 1_introduction.tex
\section{Introduction}
\label{sec:introduction}

Android's open-source nature and vast app ecosystem have contributed to its widespread adoption. For regulating access to sensitive user data, Android employs a permission framework that serves as a gatekeeper~\cite{alkindi2019android}. When using an app, the Android permission framework enables users to grant or deny permission for each type of data individually. Unfortunately, the study by Wagner et al.~\cite{ha2013android} reveals that only 17\% of Android users pay attention to permissions while installing/using an app. Once a user grants permission to an app, the permission framework allows unrestricted access to sensitive data, endangering the user's privacy~\cite{alkindi2019android}. If the user pays attention to privacy concerns and denies a permission request, the app often blocks access not only to features related to the permission but also to other features that may not require that permission~\cite{alkindi2019android}. This leaves the user with no alternative but to provide access to their data.

\begin{figure}[t]
    \centering
    \includegraphics[width=\linewidth]{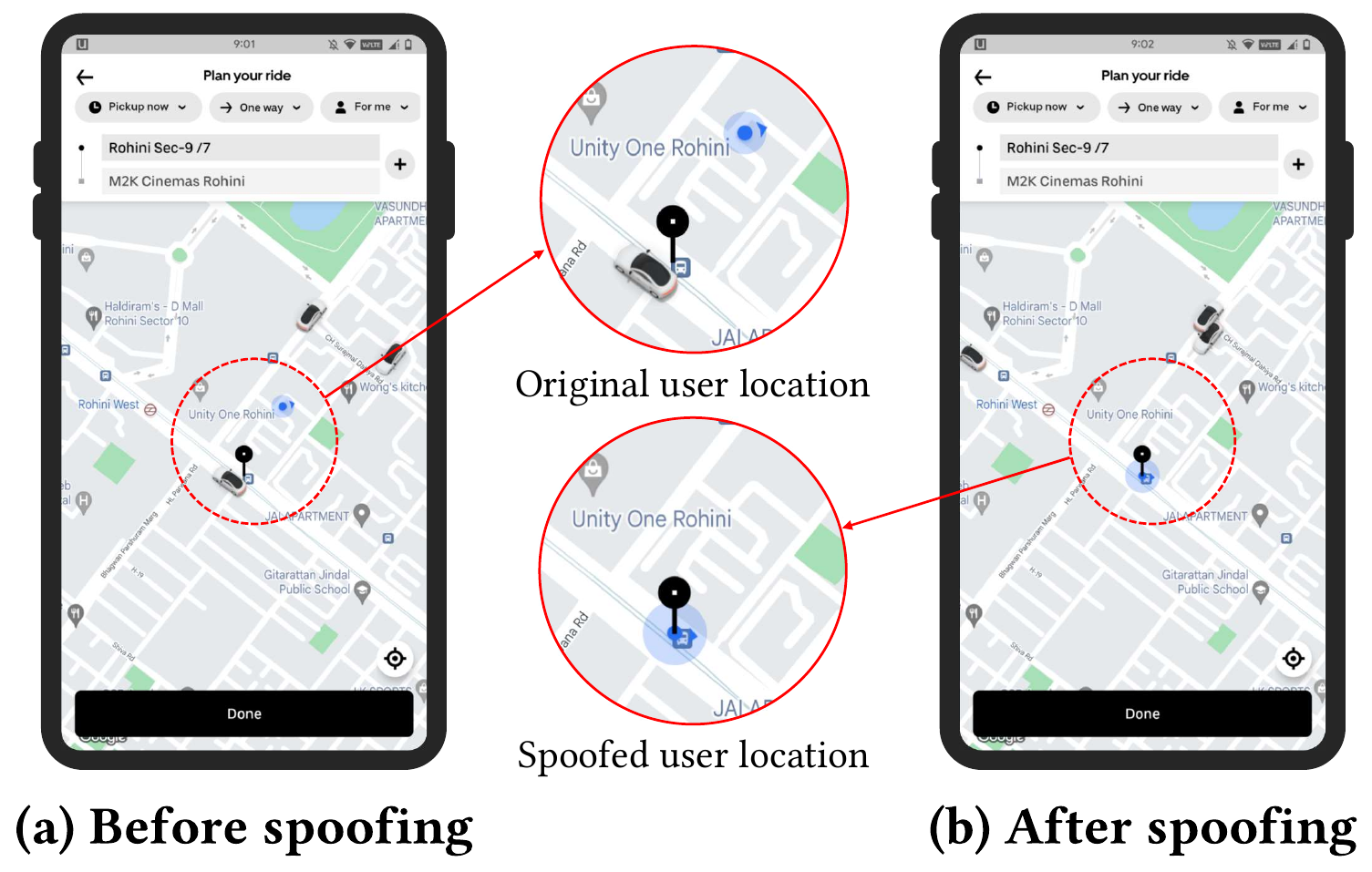}
    \caption{Screenshots illustrating how a user can share
    spoofed location using \framework{} with the Uber app.}
    \label{fig:intro-case-study-uber}
    \vspace{-10pt}
\end{figure}

To enhance user privacy without limiting the app's functionality, users could be equipped with another option: feeding user-controlled, spoofed data to apps. Figure~\ref{fig:intro-case-study-uber} shows such a case study. In Figure~\ref{fig:intro-case-study-uber}a, we observe that even though the user wants to take the cab from the pinned location, Uber encourages the user to share their original location, compromising their privacy. Figure~\ref{fig:intro-case-study-uber}b demonstrates how the user can hide their original location and feed the user-specified spoofed location as the current location. This safeguards the user's privacy while still enabling them to utilize Uber's service of finding a cab.

Existing approaches for spoofing user data typically rely on either modifying the Android OS~\cite{raval2016you, smalley2013security, wu2017context} or rebuilding the target app’s binary~\cite{backes2015boxify, jeon2012dr}. Both strategies suffer from significant usability and deployment limitations. OS-level modification requires rooting the device, a process inaccessible, insecure, and undesirable for most users~\cite{zhang2015android}. Binary rewriting, on the other hand, is fragile and increasingly ineffective: repackaged apps are readily detected by Google’s Play Integrity API~\cite{andPlayIntAPI}, often causing the app to disable its services or refuse to run.

To equip users with a practical tool for preserving their privacy, we build \framework{}, a comprehensive and robust user data spoofing system for non-rooted, stock Android devices. \framework{} intercepts permission-protected API calls at runtime and can transparently supply spoofed values for any sensitive data, including location, sensors, microphone, camera metadata, and storage access, without modifying the OS or repackaging the app. Beyond user-configured spoofing, \framework{} continuously monitors sensitive API usage to detect suspicious or privacy-violating behaviour (e.g., unexpected sensor sampling, background audio recording, covert file uploads) and automatically responds by feeding sanitized data or blocking malicious operations while preserving app functionality.

\input{tables/highlights}

\framework{} addresses two core privacy scenarios. First, in a user-driven scenario, privacy-aware users can explicitly configure spoofed values (e.g., fake location, sanitized sensor streams) to be fed to apps while retaining full functionality. Second, in a system-driven scenario, \framework{} autonomously detects suspicious behaviours such as background audio recording, covert file uploads, or sensor-based side-channel attacks—and proactively protects the user by spoofing or constraining data exfiltration, even when the user is unaware of the ongoing violation.

In our broader study of 70 Android apps, we identify four representative classes of privacy-violating behaviour that \framework{} successfully detects and mitigates by supplying spoofed data or constraining data exfiltration. A summary of representative use cases is presented in Table~\ref{tab:highlights}.

\begin{itemize}[leftmargin=*]
\item \textbf{User-driven spoofing:} When a user feels uncomfortable sharing real data, they can explicitly invoke \framework{} to supply spoofed data instead. In our case studies of Uber, Snapchat, and Truecaller, these apps request access to the user’s location, images, and text messages, respectively. \framework{} successfully injects user-configured spoofed data in each case without interrupting or degrading the apps’ normal functionality.

\item \textbf{Unexpected data usage:} In a Facebook case study, \framework{} detects the app passively recording microphone audio while the user is only scrolling the news feed. \framework{} mitigates this by supplying benign spoofed audio, preserving privacy. Prior research~\cite{awan2020violation, chapman2023exploring} has reported such audio-based privacy risks, yet users have limited control. Our demonstration video is available on \href{https://www.youtube.com/watch?v=sXiGUoqFLmk}{YouTube}\footnote{https://www.youtube.com/watch?v=sXiGUoqFLmk}. Other banned apps (Table~\ref{tab:malicious_apps}) exhibit similar behaviours, all effectively mitigated by \framework{}.

\item \textbf{Side-channel attacks:} While a user interacts with a sensitive app (e.g., a banking application), a concurrently running malicious app can sample motion and other sensors to infer sensitive information such as keyboard input or screen touch positions. We show that \framework{} can detect and mitigate such sensor-based side-channel attacks by intercepting and supplying spoofed sensor readings to untrusted apps. In a case study with the banned app Gyrosys~\cite{lin2019motion}, the touch-position prediction accuracy drops from 81.22\% to just 5.36\% when \framework{} is enabled, effectively neutralizing the attack.

\item \textbf{Protection from malicious apps:} We collectively analyse the various malicious apps (see Table~\ref{tab:malicious_apps}). \framework{} is able to detect the malicious behaviour. \framework{} is also effective against apps that directly exfiltrate private data. For example, while experimenting with the ``All Good PDF Scanner'' Android app (later banned by Google), \framework{} detected that the app was uploading users’ private files to the internet. In response, \framework{} blocked the app’s network access for the offending operations without crashing the app or disabling its benign features, thereby protecting the user’s private information in a fine-grained manner.
\end{itemize}

Across the 70 popular pre-installed and user-installed Android apps we evaluated, \framework{} successfully spoofed 78.32\% of all permission-protected data accesses while remaining fully undetected and causing no crashes or functional regressions. This demonstrates that practical and fine-grained privacy protection, which are traditionally achievable only through rooting, OS modification, or brittle app repackaging—can instead be realised on stock Android devices. Notably, \framework{} preserves full app functionality even when supplying spoofed data, mitigates a wide range of privacy-violating behaviours (including background audio capture, covert file exfiltration, and sensor-based side-channel attacks), and operates without triggering Play Integrity checks. Together, these results show that \framework{} offers a deployable, comprehensive, and user-centric privacy mechanism within the existing Android ecosystem.

In addition to analysing policy-violating behaviours of apps, we also perform an extensive evaluation of the runtime overhead introduced by \framework{}. We systematically measure its impact in both user-driven spoofing scenarios and automatic mitigation of passive privacy violations. Our results show that spoofing sensitive user data with \framework{} incurs only a minimal average overhead of 2.52\% in battery drain and 5.2 MB of additional memory during app execution. Each spoofed API call adds, on average, 1.64 ms of latency, which does not cause any noticeable performance degradation or UI lag in the apps we tested. Across all experiments, target apps continued to function normally and did not detect the presence of \framework{}, confirming that our approach is efficient for real-world deployment.

Overall, we make the following key contributions.
\begin{itemize}[leftmargin=*]
    \item We design and implement \framework{}, a comprehensive user data spoofing system for non-rooted, commodity Android devices that requires neither OS nor app-binary modification, and unifies (i) user-driven spoofing of sensitive data (e.g., location, sensors, microphone) with (ii) automatic privacy-abuse detection and response over sensitive APIs.

    \item We show that \framework{} exposes and mitigates four representative classes of real-world privacy violations, including active user spoofing, unexpected microphone recording, motion-sensor-based side-channel attacks, and direct private file exfiltration, and we demonstrate its effectiveness on widely used apps such as Facebook, Snapchat, Truecaller, Uber, Gyrosys, All Good PDF Scanner, and other policy-violating apps.

    \item We perform a comprehensive evaluation of \framework{} on 70 popular pre-installed and user-installed Android apps and show that it effectively preserves user privacy without causing crashes or being detected, while incurring only 2.52\% additional battery drain, 5.2~MB memory overhead, and 1.64~ms added latency per spoofed API call, demonstrating its practicality as a user-centric privacy mechanism within the existing Android ecosystem.
\end{itemize}

The rest of the paper is organized as follows. Section~\ref{sec:related_work} reviews prior approaches. Section~\ref{sec:motivation} motivates the need for a system capable of spoofing user data. Section~\ref{sec:methodology} describes the methodology used by \framework{} to intercept and spoof sensitive data, and Section~\ref{sec:architecture} details its system architecture. Sections~\ref{sec:mitigating_sca} and~\ref{sec:protecting_up} demonstrate how \framework{} mitigates privacy attacks from malicious apps and empowers users to control their privacy in real-world applications. Section~\ref{sec:results} presents the performance evaluation and overhead analysis, and Section~\ref{sec:conclusion} concludes the paper.

%% file: tables/highlights.tex
\begin{table*}
    {\centering
    \begin{tabular} {>{\arraybackslash}m{3cm} >{\arraybackslash}m{14cm}}
        \hline
         \textbf{Use Case} & \textbf{Example(s)}\\
         \hline
         Bypassing Continuous Authentication & \textbf{MGRA}~\cite{hong2016mgra} (Sec \ref{sec:continuous_authentication_mechanisms}) utilizes the accelerometer data and \textbf{VoiceLive}~\cite{zhang2016voicelive} (Sec \ref{sec:continuous_authentication_mechanisms}) utilizes the smartphone's two different microphones to authenticate users. But \framework{} proves to be capable of bypassing these authentication mechanisms by spoofing the input data. \\
         \hline
         Protection from Malicious Apps & \textbf{All Good PDF Scanner} (Sec \ref{sec:malicious_apps}) was granted permissions like Storage and Camera but was detected to be uploading data on the internet in the background, hence its internet access was blocked by \framework{}. \textbf{Unique Keyboard} (Sec \ref{sec:malicious_apps}) was found to be accessing the sensor data while running in the background, therefore \framework{} fed deceived data when app is running in the background.\\
         \hline
         Side-Channel Attack Mitigation & \textbf{Gyrosec}~\cite{lin2019motion} (Sec \ref{sec:side_channel_attack}) records the sensor data while running in the background, and uploads it over the internet. This was mitigated by deceiving the sensor data received by the apps.  \\
         \hline
         Unexpected Sensor Usage Detection & \textbf{Facebook} (Sec \ref{sec:fb_case_study}) requested for Audio permission. But it was found (using \framework{}) to be recording the microphone data when no activity requiring microphone was being used by the user.\\
         \hline
         User Privacy Protection & \textbf{Snapchat} (Sec \ref{sec:sc_case_study}) and \textbf{Truecaller} (Sec \ref{sec:tc_case_study}) provide great features to users but at the cost of sharing private information like Location, Contacts and Messages. \framework{} enables users to enjoy the provided features without compromising private information.\\
         \hline
    \end{tabular}
    }
    \caption{Various use cases of \framework{} and their respective real-world example(s)}
    \label{tab:highlights}
\end{table*}

%% file: 8_related_work.tex
\section{Related Work}
\label{sec:related_work}


Mobile privacy has been addressed through several classes of approaches, including modifying the OS or app code~\cite{backes2015boxify, wu2017context}, revoking permissions~\cite{chen2013contextual, chen2017sweetdroid}, completely or selectively blocking permission data~\cite{bokhorst2017xprivacy, ratazzi2019pinpoint, hornyack2011these, shrestha2016slogger}, rule-based or context-aware data filtering~\cite{bokhorst2021xprivacylua, raval2019permissions}, and permission-usage labelling mechanisms~\cite{chitkara2017does, tsai2017turtle}. Each category provides partial protections with different trade-offs in usability, functionality, and deployment practicality. Table~\ref{tab:deceiverRelatedWorkComparison} summarises these approaches and contrasts them with the comprehensive capabilities offered by \framework{}.

\input{tables/related_work_comparison}

Several systems control or spoof data access by modifying either the Android OS~\cite{backes2015boxify, wu2017context} or the target app binaries~\cite{jeon2012dr, raval2016you}. For example, Boxify~\cite{backes2015boxify} intercepts Binder calls by altering the OS, and BinderFilter~\cite{wu2017context} enforces contextual constraints by modifying system components. App-rewriting solutions such as MrHide~\cite{jeon2012dr} and PrivateEye~\cite{raval2016you} rebuild app binaries to interpose on sensitive API calls. Although effective in controlled environments, these techniques require rooting, system-level modifications, or repackaging apps—steps that are impractical for average users and often lead to crashes, incompatibility with Play Store apps.

Approaches such as Pegasus~\cite{chen2013contextual} and SweetDroid~\cite{chen2017sweetdroid} dynamically revoke permissions based on contextual triggers or user-defined policies. While revocation provides fine-grained control, many apps assume granted permissions remain available and crash or disable core features when permissions are revoked. As a result, revocation-based systems often compromise usability and do not offer a seamless privacy-preserving experience.

Complete data-blocking solutions such as XPrivacy~\cite{bokhorst2017xprivacy} and PINPOINT~\cite{ratazzi2019pinpoint} prevent apps from accessing sensitive data altogether, whereas selective blocking approaches like AppFence~\cite{hornyack2011these} and Slogger~\cite{shrestha2016slogger} attempt to suppress only specific data flows. Rule-based data filtering (e.g., XPrivacyLua~\cite{bokhorst2021xprivacylua}) and context-aware filtering mechanisms~\cite{raval2019permissions} extend this idea by shaping data access based on policies or runtime context. However, these systems still disrupt normal app behavior, do not deceive apps into accepting fabricated data, and may break functionality when essential data is suppressed.

Systems like ProtectMyPrivacy~\cite{chitkara2017does} and TurtleGuard~\cite{tsai2017turtle} notify users when apps access sensitive resources, improving transparency but not preventing privacy violations once permission is granted. These tools lack the ability to block, filter, or spoof data, limiting their effectiveness when apps behave maliciously or unexpectedly.

As summarised in Table~\ref{tab:deceiverRelatedWorkComparison}, existing approaches each address only a subset of the privacy problem. Many require source code modification, revoke permissions that break app functionality, or fail to deceive apps—making them unsuitable for everyday users on stock Android devices. Critically, no prior work simultaneously offers: (i) no OS or binary modification, (ii) no permission revocation, (iii) support for blocking, filtering, prompting, and spoofing permission data, (iv) preservation of app functionality, and (v) operation on non-rooted devices. In contrast, \framework{} integrates all these capabilities within a single deployable system. Additionally, it supports user-driven spoofing as well as automatic detection and mitigation of privacy-violating behaviours. This unique combination makes \framework{} a practical, comprehensive, and real-world-deployable mobile privacy solution.

%% file: tables/related_work_comparison.tex
\begin{table*}[htbp]
\caption{Comparison of \framework{} with related approaches based on supported capabilities}
\begin{center}
\begin{tabular}{|M{2.6cm}|M{1.6cm}|M{1.2cm}|M{1.0cm}|M{1.0cm}|M{1.4cm}|M{1.2cm}|M{1.3cm}|M{1.3cm}|M{1.3cm}|}
\hline
\textbf{Approach} & \textbf{No source modification} & \textbf{Not Revokes Perm.} & \textbf{Blocks Perm. Data} & \textbf{Filters Perm. Data} & \textbf{Prompts on Perm. Use} & \textbf{Deceives Perm. Data} & \textbf{Works without Rooting} & \textbf{User Driven Control} & \textbf{Preserve App Func.} \\
\hline
OS/App Code Mods. ~\cite{backes2015boxify, wu2017context} & \xmark & \cmark & \cmark & \cmark & \xmark & \xmark & \xmark & \xmark & \xmark \\
\hline
Revoking Perm.~\cite{chen2013contextual, chen2017sweetdroid} & \cmark & \xmark & \cmark & \xmark & \xmark & \xmark & \cmark & \cmark & \xmark \\
\hline
Complete Data Blocking ~\cite{bokhorst2017xprivacy, ratazzi2019pinpoint} & \cmark & \cmark & \cmark & \xmark & \xmark & \xmark & \xmark & \xmark & \xmark \\
\hline
Selective Data Blocking~\cite{hornyack2011these, shrestha2016slogger} & \xmark & \cmark & \cmark & \cmark & \xmark & \xmark & \xmark & \cmark & \xmark \\
\hline
Rule-based Data Filtering~\cite{bokhorst2021xprivacylua} & \cmark & \cmark & \cmark & \cmark & \cmark & \xmark & \xmark & \cmark & \xmark \\
\hline
Context-aware Data Filtering~\cite{raval2019permissions} & \cmark & \cmark & \cmark & \cmark & \cmark & \xmark & \cmark & \cmark & \xmark \\
\hline
Labelling Perm. Usage~\cite{chitkara2017does, tsai2017turtle} & \cmark & \cmark & \xmark & \xmark & \cmark & \xmark & \cmark & \cmark & \cmark \\
\hline
Proposed Frame- work (\framework{}) & \cmark & \cmark & \cmark & \cmark & \cmark & \cmark & \cmark & \cmark & \cmark \\
\hline
\end{tabular}
\label{tab:deceiverRelatedWorkComparison}
\end{center}
\end{table*}

%% file: 2_motivation.tex
\section{Motivation}
\label{sec:motivation}

The motivation for our work stems from the realization that current approaches for protecting user data in the Android ecosystem are insufficient leaving user data exposed to privacy violations from the installed apps. 

\subsection{Inadequate Android Permission Framework}

The Android platform provides a permission framework that enables the user to control the data access privileges granted to individual apps. It aims to safeguard sensitive user data by restricting unauthorized access. However, it has been found to be inadequate failing to prevent apps from collecting sensitive data. Researchers~\cite{hasan2013sensing, simon2013pin, ba2020learning, shen2015input, lin2019motion} have shown that non-runtime permissions, also referred to as ``normal permissions'', like those related to the inertial measurement unit, can contain sensitive information, but the Android permission framework grants access to them without notifying the user.

For runtime permissions, also referred to as ``dangerous permissions'', like location, requested by apps, the user is presented with a binary choice to either grant or deny each permission. Granting a requested permission can result in a privacy breach while denying it triggers a \texttt{SecurityException} in the app. If the exception is not handled properly, the app crashes. Even if the exception is handled correctly, due to a lack of permission, the app limits its services. Further, Wagner et al.\cite{wijesekera2017feasibility} noted that only 17\% of users pay attention to permissions during installation. Recently added options, such as ``\textit{Only this time}'' and ``\textit{While using the app}'', in the permission framework aim to protect user privacy by preventing background permission usage. Hence, the existing permission framework lacks flexibility, leaving users with limited choices.

\subsection{Limitations of Existing User Data Spoofing Mechanisms}

The aforementioned inadequacy of the Android permission framework has motivated researchers to look for an alternative option, i.e., to deceive the user data fed to the apps. While modifying the source code of the Android OS or target apps~\cite{backes2015boxify, jeon2012dr, raval2016you, smalley2013security, wu2017context} might seem like a solution, this approach necessitates the creation of a custom ROM. However, installing custom ROMs on Android devices requires root privileges, making it impractical for most users. Additionally, modifying the source code of each target app is not a scalable solution. This mechanism can also be easily detected if Google's Play Integrity API is utilized in the target app~\cite{andPlayIntAPI}.

\input{tables/various_continuous_authentication_approaches_and_deceivable_status}

\subsection{Achilles Heel of Continuous User Authentication}
\label{sec:continuous_authentication_mechanisms}

The conventional knowledge-based authentication approaches require a user to provide information like passwords to access their device. While these methods are simple to implement, they suffer from drawbacks such as the need for frequent re-entering and the possibility of reuse by an attacker. To mitigate these limitations, continuous authentication mechanisms have been implemented that verify the user's identity based on their behavioral biometrics such as keystroke dynamics, touch gestures, motion, and voice. By leveraging such inherent biometric signatures of the user, this approach aims to prevent unauthorized access. Table \ref{tab:ca_approaches_deceivable} offers an overview of different continuous authentication mechanisms across five widely used modalities.

The user's biometric signatures are captured implicitly through various data streams, including interaction, environmental information, and sensory data. Hence, the underlying assumption to justify the security of these mechanisms is that these biometric signatures are unique to the user and difficult for attackers to replicate. Upon analyzing these approaches, we determined that the majority of them rely on user data (specifically, sensor data) as the primary data source for user authentication. Hence, these authentication mechanisms can be bypassed by manipulating the user data (especially sensor data) fed into them. The attacker can simply present false sensor data replicating the data recorded when the authentic user was utilizing the device. This can lead the authentication mechanism to erroneously conclude that the genuine user is using the device and allow the attacker to gain unauthorized access to sensitive data or perform actions on behalf of the legitimate user. 

However, we note that since there is no robust mechanism for spoofing the user data in the existing literature, the aforementioned attack on the continuous authentication mechanisms is difficult to realize in the real world. The absence of such a strong attack has resulted in the wide acceptance of weak continuous authentication methods~\cite{kolokas2019gait, sun2018artificial, thang2012gait, hoang2013adaptive, shih2015flick, nohara2016personal, lu2015safeguard, jain2015exploring, nixon2016slowmo, feng2014tips, abuhamad2020autosen, amini2018deepauth, li2018using, yan2018towards, song2016eyeveri, xia2018motionhacker, hong2016mgra, hong2015waving, miguel2016interaction, zhang2016voicelive, wang2019voicepop, johnson2013secure, khamis2016gazetouchpass, zhu2013sensec, sitova2015hmog, pang2019mineauth, acien2019multilock, zhu2019riskcog, lee2017implicit}.

To mitigate the insufficiency of the Android permission framework and to establish a practical benchmark for continuous authentication mechanisms, there is a need for a more robust and comprehensive approach to spoof user data in the Android ecosystem. This paper aims to address this critical gap by identifying and addressing the limitations of existing data spoofing mechanisms.

%% file: tables/various_continuous_authentication_approaches_and_deceivable_status.tex
\begin{table}
    {\centering
    \begin{tabular}{>{\centering\arraybackslash}m{1.8cm} > {\centering\arraybackslash}m{1.7cm} >{\centering\arraybackslash}m{2.1cm} >{\centering\arraybackslash}m{1.5cm}}
        \hline
          \textbf{Modality} & \textbf{Framework} & \textbf{User Data} & \textbf{Accuracy} \\
         \hline
         \textbf{Gait} \\
         & ~\cite{kolokas2019gait, sun2018artificial, thang2012gait, hoang2013adaptive} & Ac, Ca, Gy, Mg & 91-95\% \\
         \hline
         
         \textbf{Gesture} \\
         Flick & ~\cite{shih2015flick, nohara2016personal} & Ac, Gy & 92.8-98\% \\
         
         Swipe & ~\cite{lu2015safeguard, jain2015exploring} & Ac, Or, To \\
         
         Touch & ~\cite{nixon2016slowmo, feng2014tips} & To & 89\%\\
         \hline

         \textbf{Motion} \\
         Free & ~\cite{abuhamad2020autosen, amini2018deepauth, li2018using} & Ac, Gy, Mg & 96.7-97.5\% \\
         
         Shake & ~\cite{yan2018towards} & Ac, Gy & 96.87\%\\

         Eye & ~\cite{song2016eyeveri} & Ca & 88.73\% \\

         IaH & ~\cite{xia2018motionhacker} & Ac, Gy & 32.8\% \\
         
         Gesture & ~\cite{hong2016mgra, hong2015waving} & Ac & 92.2-95.8\%\\

         \hline

         \textbf{Voice} \\
         & ~\cite{miguel2016interaction, zhang2016voicelive, wang2019voicepop, johnson2013secure} & Mi & 93.5-99.3\% \\
         \hline 
        
         \textbf{Multimodal} \\
         Ge, Mo & ~\cite{khamis2016gazetouchpass, zhu2013sensec} & Ac, Gy, Or, To & 65\% \\
         
         Ga, Ge, Mo & ~\cite{sitova2015hmog} & Ac, Gy, To & \\

         Be, Ga, Ge & ~\cite{pang2019mineauth, acien2019multilock} & Ac, Nw, Tk, To & 85-97.1\% \\
         
         Ga, Ge & ~\cite{zhu2019riskcog, lee2017implicit} & Ac, Gy, Li, Mg & 95.6-98.1\% \\

        \hline
        \multicolumn{4}{p{7.2cm}}{\footnotesize \textbf{Ac}: Accelerometer, \textbf{Ca}: Camera,  \textbf{Gr}: Gravity Sensor, \textbf{Gy}: Gyroscope, \textbf{Li}: Light Sensor, \textbf{Mg}: Magnetometer,  \textbf{Mi}: Microphone, \textbf{Nw}: Network, \textbf{Or}: Orientation, \textbf{To}: Touch, \textbf{Tr}: Tracking} \\
        \hline
        
        \multicolumn{4}{p{7.2cm}}{\footnotesize \textbf{Be}: Behaviour, \textbf{Ga}: Gait, \textbf{Ge}: Gesture, \textbf{IaH}: In-air Handwriting, \textbf{Mo}: Motion}\\
        \hline
    \end{tabular}
    }
    \caption{Existing continuous authentication methods.}
    \label{tab:ca_approaches_deceivable}
\end{table}

%% file: 3_methodology.tex
\section{\framework{} Methodology}
\label{sec:methodology}

We now provide the methodology to achieve the goal of deceiving user data fed to the apps without rooting the device or altering the source of the apps.

\subsection{Background on Hooking}
In Android, a typical user does not have permission to modify the apps or OS behaviour. However, \textit{hooking} into OS and apps is one of the most practical options to alter the behaviour of an app. Hooking is a technique of code modification that changes the behaviour of the original code running sequence by inserting instructions into the code segment. 

Hooking can be performed in two ways: statically and dynamically. The static hooking technique injects hooks before app execution by altering the byte code of the app or by injecting custom shared libraries. However, static hooking causes permanent alteration and can be detected by an app using a digital signature. 

Dynamic hooking technique applies ``hooks'' at runtime by modifying the code stored in volatile memory (temporary modification), allowing hooking decisions based on the runtime environment. Android's Dalvik VM utilizes Dynamic Linkers' ``virtual method table'' to jump to the memory location on a method call, hence \textit{Call Diversion} can be employed to inject hooks in the hooked method lookup table entry. 

\begin{figure}[t]
\centering
\begin{subfigure}{0.9\linewidth}
    \includegraphics[width=\linewidth]{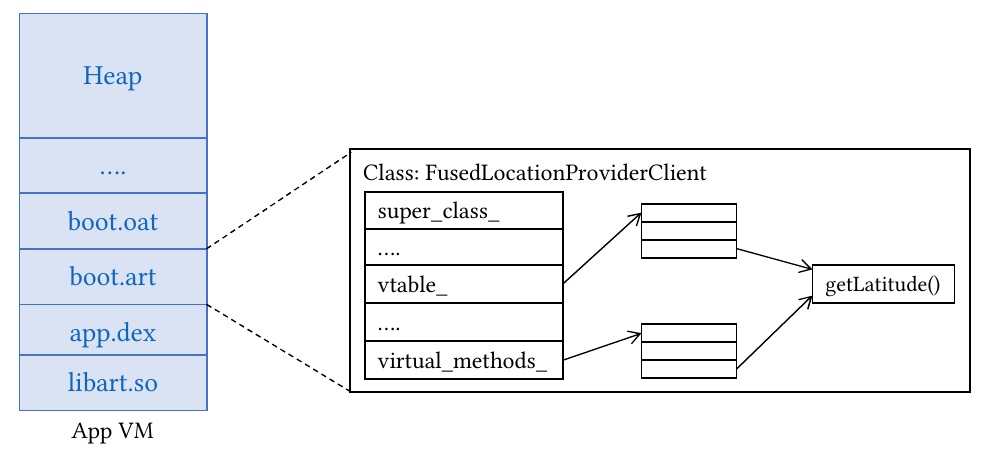}
    \caption{Original method call.}
    \label{fig:vrtulMemMthdCall_woFr}
\end{subfigure}
\begin{subfigure}{0.9\linewidth}
    \includegraphics[width=\linewidth]{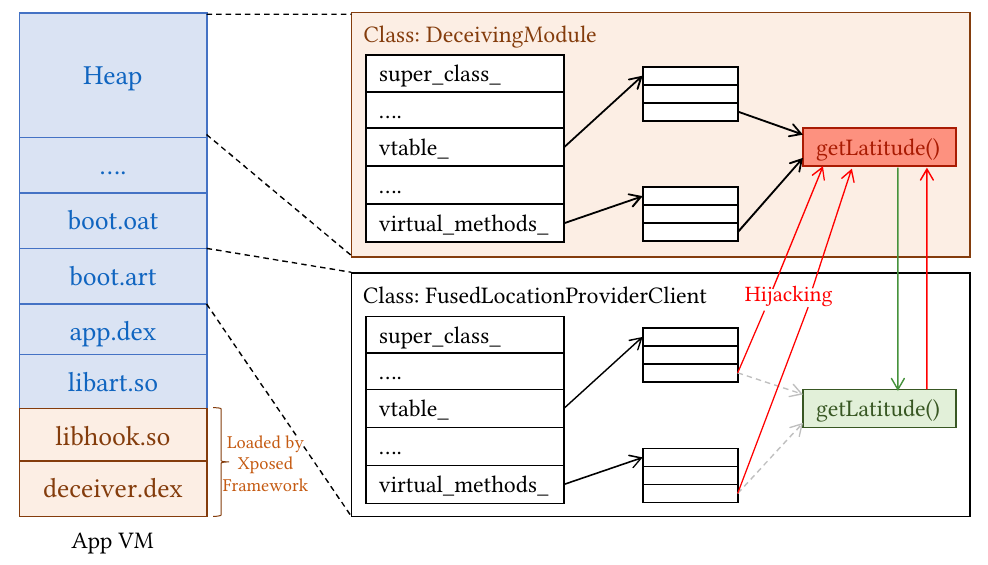}
    \caption{Method call with \framework{}.}
    \label{fig:vrtulMemMthdCall_wFr}
\end{subfigure}
\caption{Example illustrating how the Xposed module class method hijacks the original method call.}
\label{fig:vrtulMemMthdCall}
\end{figure}

In Figure \ref{fig:vrtulMemMthdCall}, we present an overview of dynamic hooking in Android using the \texttt{Location.getLatitude()} method. Figure \ref{fig:vrtulMemMthdCall_woFr} illustrates the conventional process of method call jumps. The virtual memory of the target app encompasses the heap space and various files loaded by the \textit{Zygote} process to initiate the app process. Zygote is a process started by the Android Framework during the OS boot. It is a template process preloaded with Java-shared libraries in memory, responsible for launching apps and services. 

The \texttt{boot.oat} and \texttt{boot.art} files serve as boot images to speed up the boot proces containing frequently used class source code and the pre-initialized heap respectively. The heap of the target class is stored in the \texttt{boot.art} file. The Dalvik Executable code of the app and the shared libraries are stored in the \texttt{app.dex} and \texttt{libart.so} respectively. When the target method is invoked by the app, the existence of the target class is checked in the \texttt{boot.art} file. If the class is found, the addresses of the target methods are determined using the virtual tables stored in the heap. Consequently, the execution jumps to the identified address.

Figure \ref{fig:vrtulMemMthdCall_wFr} depicts the process of hooking. The hooked application's virtual memory is loaded with the \texttt{libhook.so} shared library by the hooking module, facilitating the hooking operations. Subsequently, the hooks defined by the hooking app (\texttt{hook.dex}) are loaded by the hooking module as Dalvik Executables (same as app code). Upon completion, \texttt{libhook.so} intervenes and loads the hooking classes into the app's heap space, modifying the entries of the target class virtual tables with the addresses of the hooking class methods. Consequently, when the app calls the hooked methods, the calls are redirected to the hooking class methods instead of the original methods. Finally, the original method is invoked by storing its address at the end of the hooking method.

\input{listings/deceiving_mechanism}

\subsection{Hooking on Non-Rooted Device}

We found that the most suitable solution to deceive user data is app hooking. However, on non-rooted devices, users are restricted from accessing the root and hooking various processes within the system. To gain hooking capabilities, root access to the \textit{Zygote} process is required. 

For this, we utilize \textit{adb} (\textit{Android Debug Bridge}), a command-line tool typically utilised for debugging apps, but also capable of controlling Android devices and executing shell commands. \textit{adb} provides user-installed apps with privileged access to the system APIs. This access is limited till the device is in an active connection with the \textit{adb} tool. This limitation is futher tackled by the Shizuku service by exploiting \textit{adb} shell access. Shizuku provides privileged access to system APIs to user-installed apps by running a dedicated process with shell-level permissions, acting as a proxy between the apps and the OS. This level of access is sufficient for hooking the target apps to spoof the user data. 

Employing the Shizuku service, we use LSPatch, an Xposed framework for non-rooted devices, for hooking the target apps by injecting the hooks into app processes. This allows users to define the required app behaviour within the Xposed module to make the target app execute their desired actions. Both the Shizuku service and LSPatch are open-source and user-controlled, they only grant root privileges to the modules and apps chosen by the user, ensuring that the attack surface remains unchanged.

\subsection{Spoofing User Data with Hooking}

In Listing \ref{lst:clipboardDeceiver}, we present the approach employed to spoof the Clipboard data with custom spoofed data by hooking. Typically, the app retrieves clipboard data by calling the \texttt{ClipboardManager.getPrimaryClip()} method. However, to modify the result of this method, multiple other methods need to be hooked. Alternatively, a more efficient approach involves hooking \texttt{ClipData.CREATOR.createFromParcel()} method, as ultimately, this method is called by the former method to return the clipboard data. To hook this method, we utilise Xposed's \texttt{handleLoadPackage()} in Line~7.

Using this mechanism, the Clipboard data can either be blocked or modified before feeding into the target apps. In the before hook, Line 11-15 lists the code for blocking the clipboard data from feeding into the target apps. Using \textit{ContentProvider} the Xposed module is gathering the user's choice on blocking the clipboard data in Line 11-13. And then in Line 15, if the user opts for blocking the clipboard data, then the resultant part of the function call is updated with null object. This blocks the original method call and execution returns back to the method from where the call for the hooked method is made. 

Similarly, in the after hook, i.e. Lines 20-26, the clipboard data is deceived as plain text data as per the user policy. The user-defined deceived data is fetched from the \textit{ContentProvider} in Line 21-24. And then in Line 26, the result of the method call is updated with a new \texttt{ClipData} object initialised with fetched deceived result.

\framework{} offers diverse policies for deceiving various permissions. For instance, one policy for spoofing camera permission involves blurring the image when the app is operating in the background. In this scenario, \framework{} utilizes the \texttt{ActivityManager} to determine whether the target app is running in the background or not. If it is found to be running in the background while accessing the camera permission, \framework{} blurs the frames supplied to the app by manipulating the pixels. Another intriguing policy employed by \framework{} involves providing a noise audio fed if an app unnecessarily requests Audio permission.

\subsection{\framework{} and its Robustness}
In this section, we present \framework{}'s mechanism to achieve the functionality of spoofing the user data fed into other apps and discuss its performance against real-world apps highlighting its robust nature.

\subsubsection{Robust Mechanism} 

By utilising dynamic hooking, \framework{} effectively manipulates user data in other apps by modifying the source code stored in volatile memory, thus avoiding permanent alterations on disk. As a result, the injected hooks cannot be detected using APK signatures or Google's Play Integrity API, making \framework{} resilient against these security measures. Moreover, \framework{} follows a methodology of spoofing user data instead of outright revoking permissions based on context. This approach ensures that \framework{} remains crash-proof, providing a reliable and robust solution.

\subsubsection{Spoofing Real-world Apps}

The capability of \framework{} against real-world apps in protecting user privacy by spoofing was explored using 50 real-world apps downloaded from the Google Play Store, and 20 pre-installed Android apps. All permissions requested by the apps were granted beforehand, and the apps were subjected to 20 minutes of manual usage. All experiments were conducted on a \textit{Samsung Galaxy M21} smartphone with a 2.3 GHz octa-core processor and 4 GB of RAM running Android 12. The outcomes were documented for each permission, taking into account visual confirmation of spoofed user data and the logs captured by \framework{}.

\begin{figure}[t]
\includegraphics[width=\linewidth]{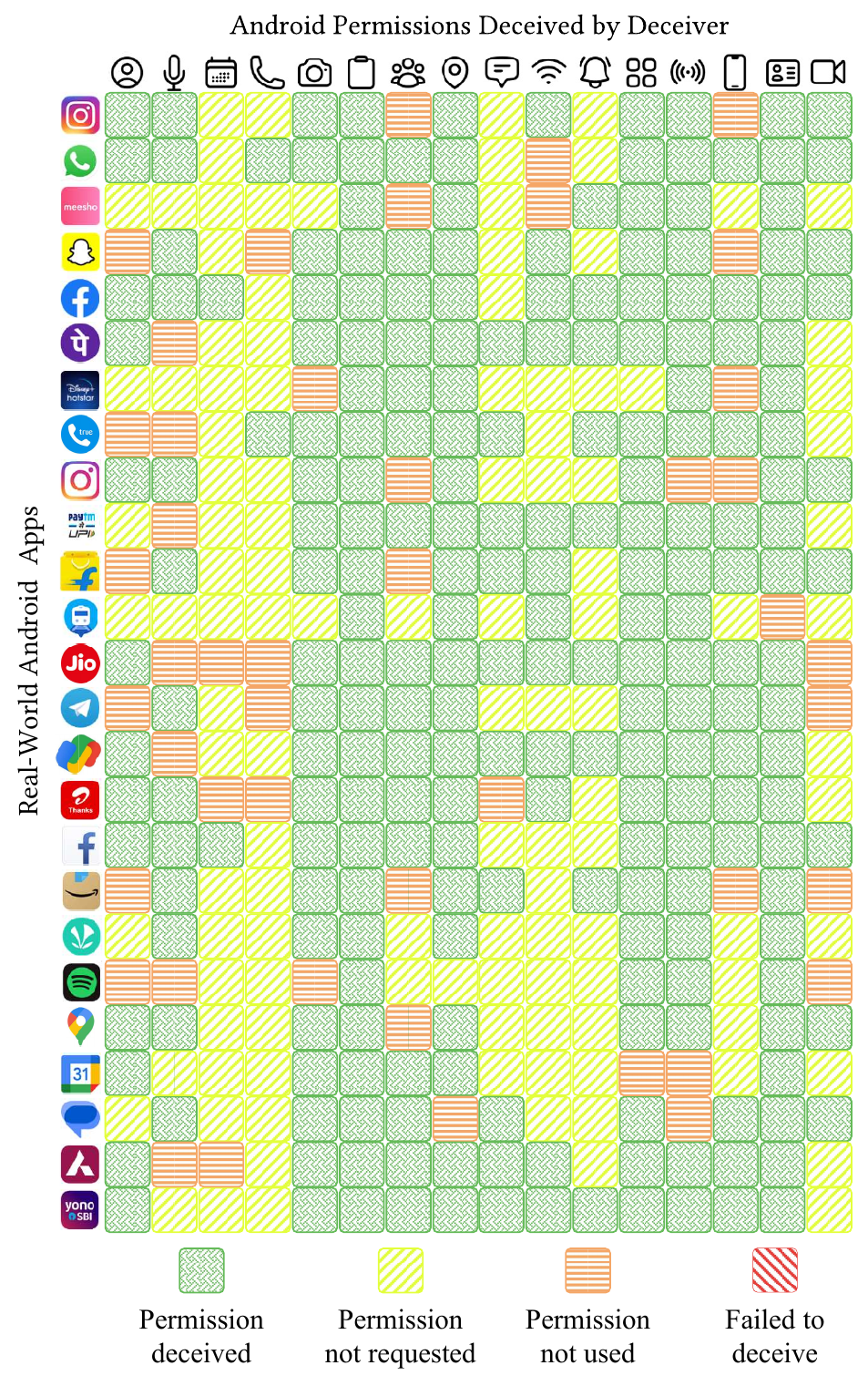}
\caption{Heatmap illustrating status of user data deceived for various popular apps by \framework{}.}
\label{fig:intro_heatmap}
\end{figure}

The heatmap depicted in Figure \ref{fig:intro_heatmap} provides an overview of the \framework{}'s performance when applied to the most popular apps on the Google Play Store. The dark green region of the heatmap signifies the successful deception of permissions for corresponding Android apps. The light green, orange, and red regions indicate permissions that were not requested, not utilized, and could not be deceived, respectively. 



Out of the 678 permissions requested by 70 apps, the \framework{} achieved a \textbf{78.32\%} success rate in spoofing user data. We could not ascertain the deception for 21.68\% of the permissions because although they were granted to the app, they were not explicitly requested during the experiments.
These encouraging results demonstrate the robustness of the \framework{}'s mechanism and design in real-world scenarios as none of the apps crashed and functioned as expected.

\subsubsection{Hooking Hidden APIs} Certain permissions such as sensors, involve method calls returning instances of \textit{Android Non-SDK Interfaces} classes also known as \textit{Hidden APIs} like \texttt{InputSensorInfo}. 
Accessing these Hidden APIs is restricted for \textit{JNI} and \textit{Java Reflections}. \framework{}'s hooks corresponding to Hidden APIs bypass these restrictions and instantiate classes with deceived data using the \textit{LSPlant}. Out of 78.32\% successfully deceived user data, 23.16\% were deceived by hooking hidden APIs.


\subsubsection{Isolation} \framework{} requests the \textit{Query All Packages} permission to facilitate and simplify its privacy protection functionalities which is a non-runtime (normal) permission and the only permission requested by \framework{}. Notably, \framework{} operates offline, severing any external connections and isolating itself from the internet, establishing a secure and trustworthy environment.


%% file: listings/deceiving_mechanism.tex
\begin{lstlisting}[caption={Kotlin code to deceive Clipboard permission data with the data defined by user in \textit{Deceit} policies.},label={lst:clipboardDeceiver},language=Kotlin,float=*]
class SpoofingModule: IXposedHookLoadPackage {
    companion object {
        fun getClass(clsName: String): Class<*> = Class.forName(clsName, false, lpparam.classLoader)    
    }
    
    @Throws(Throwable::class)
    override fun handleLoadPackage(lpparam: XC_LoadPackage.LoadPackageParam) {
        hookAllMethods(getClass(ClipData.CREATOR.javaClass.name), "createFromParcel", XC_MethodHook() {
            @Throws(Throwable::class)
            override fun beforeHookedMethod(param: MethodHookParam) {
                val uri = Uri.parse("content://com.xposedModule.provider/deceitSettings")
                val cursor = context.contentResolver.query(uri, null, null, null, null) ?: return
                val blockClipboard = cursor.getInt(cursor.getColumnIndex("blockClipboard")
                cursor.close()
                if(blockClipboard) param.result = null
            }

            @Throws(Throwable::class)
            override fun afterHookedMethod(param: MethodHookParam) {
                param.result ?: return
                val uri = Uri.parse("content://com.xposedModule.provider/deceits")
                val cursor = context.contentResolver.query(uri, null, null, null, null) ?: return
                val clipboardLabel = cursor.getString(getColumnIndex("clipboardLabel")) ?: "dummyLabel"
                val clipboardText = cursor.getString(getColumnIndex("clipboardText")) ?: "dummyText"  
                cursor.close()
                param.result = ClipData.newPlainText(clipboardLabel, clipboardText)
            }       
        })   
    }
}
\end{lstlisting}

%% file: 4_architecture.tex
\section{Implementation}
\label{sec:architecture}

\begin{figure}[t]
\includegraphics[width=\linewidth]{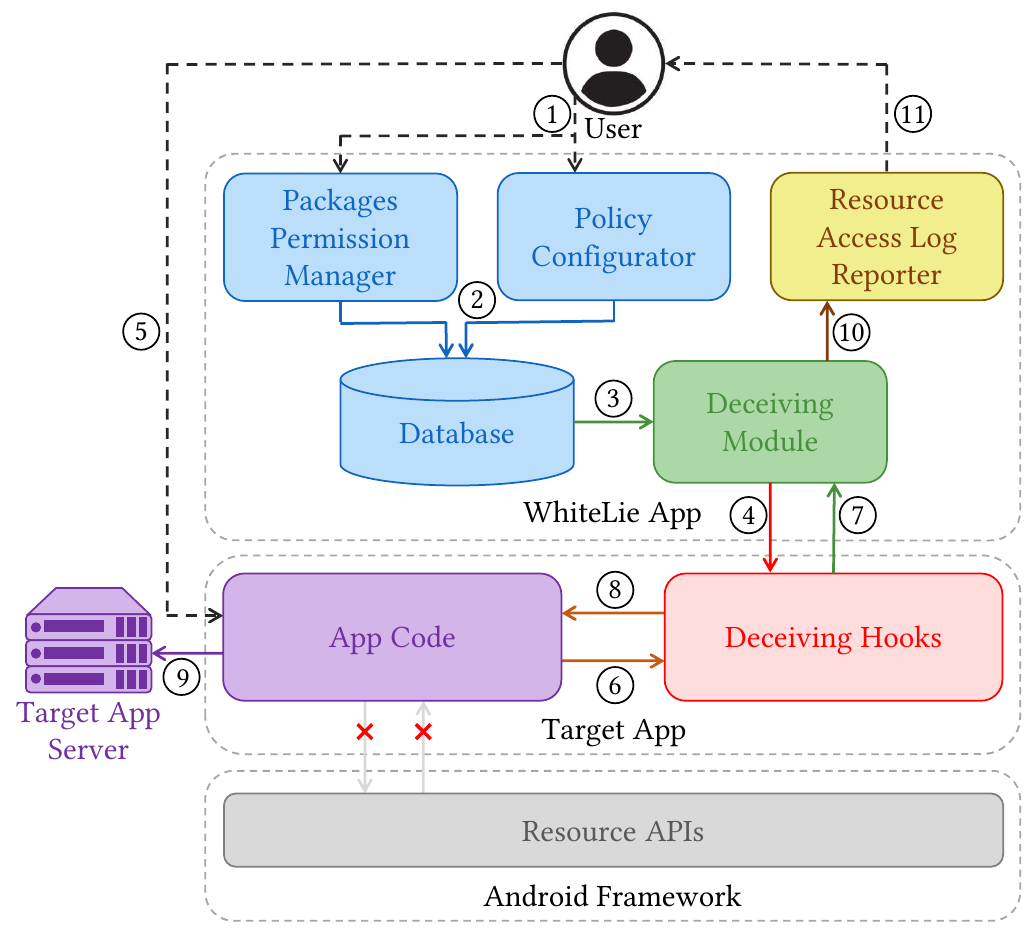}
\caption{\framework{} architecture.}
\label{fig:method_frmwrkArch}
\end{figure}

\framework{} comprises of four essential components: the \textit{Packages Permission Manager}, \textit{Policy Configurator}, \textit{Resource Access Log Reporter}, and \textit{Deceiving Module}. An overview of \framework{}'s architecture is illustrated in Figure \ref{fig:method_frmwrkArch}. \framework{} accomplishes its tasks through a series of essential functions: listing installed packages on the device along with their requested permissions, configuring user-defined policies to deceive targeted apps, hooking into the processes of selected packages, and providing detailed resource access reports for manual investigation and action.

\begin{figure}[t]
    \centering
    \includegraphics[width=\linewidth]{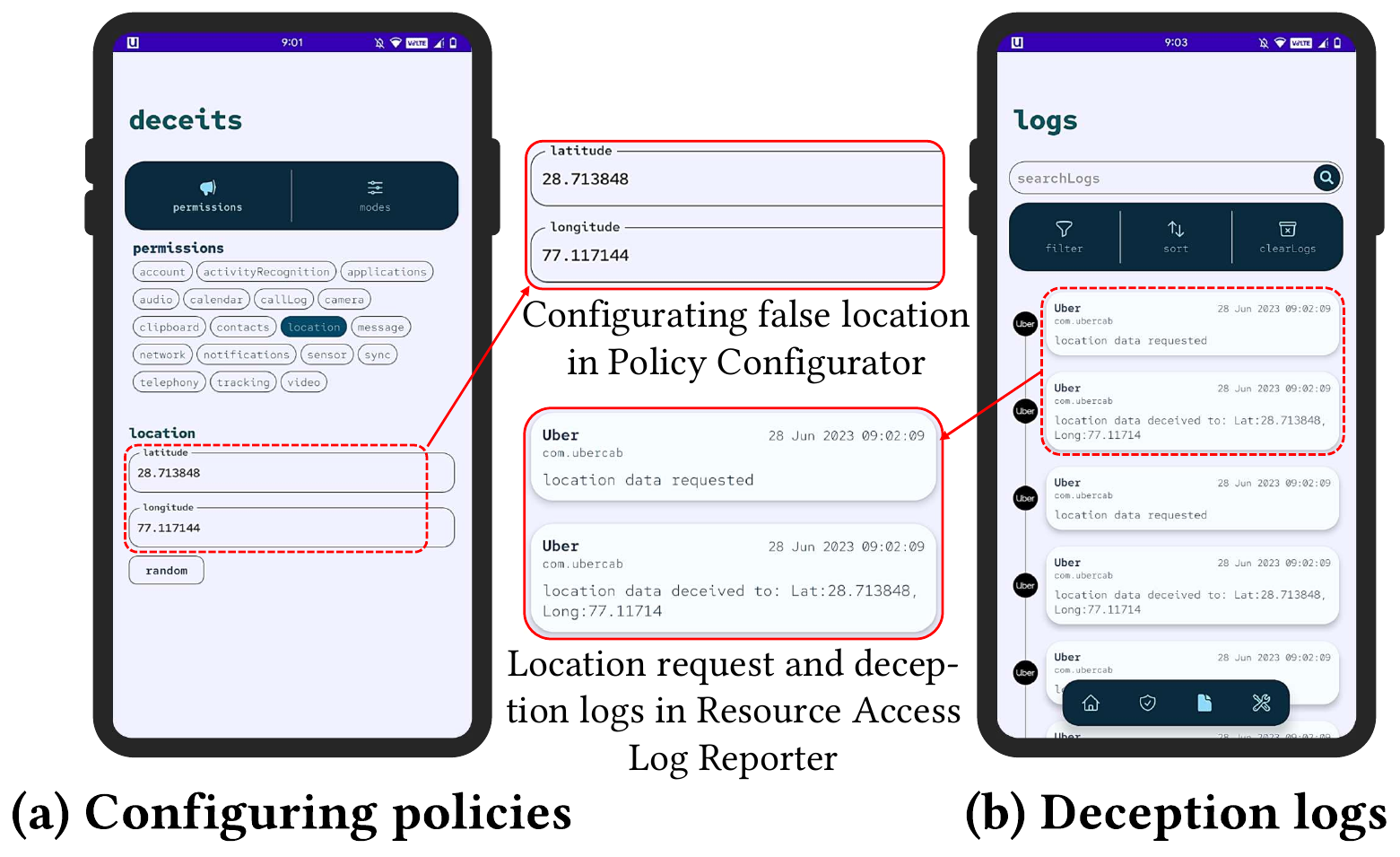}
    \caption{Screenshots of \framework{} app UI while performing experiments to deceive location data for Uber.}
    \label{fig:frmwrk-ui}
\end{figure}

\circled{1}~To deceive the user data in the target apps, the user interacts with the \textit{Package-Permission Manager} to review the permissions requested and granted to the target package, listed by \framework{} using \textit{Query All Packages} permission (sole permission requested by \framework{}). The installed packages are fetched using \texttt{getInstalledApplication()} method configured for gathering metadata of installed applications. Then the user creates policies for the deceiving the user data associated with each permission. Figure~\ref{fig:frmwrk-ui}a presents the UI of \framework{}'s \textit{Policy Configurator} while defining the new deceit data for spoofing the location permission. Configuring policies enhances user's control over the spoofed data fed to apps by customizing the data, avoiding potential privacy compromises. 

\framework{} also streamlines the configuration of privacy policies by providing predefined 
yet configurable set of meaningful real-world values and randomly selects to efficiently safeguard user data. For example, for Contacts permission, \framework{} has a set of 100 predefined real-world contacts, that are randomly fed to the target app when requested. \circled{2}~The \textit{Package-Permission Manager} and the \textit{Policy Configurator} stores the metadata and policies defined by the user in the database.




When the target app is launched and its process is created, the \textit{XposedBridge} class is loaded along side in the process which makes the active Xposed modules, including the \framework{} to interact with the app.
\circled{3}~The \textit{Deceiving Module} of \framework{} fetches the policies defined by the user from the database and \circled{4}~accordingly installs the hooks in the target app.

\circled{5}~When the user interacts with the target app, \circled{6}~the app invokes the \textit{Deceiving Hooks} methods, instead of invoking the original sensitive Resource APIs. \circled{7}~These methods communicate with the \textit{Deceiving Module} via \texttt{ContentProvider} which references the stored policies to make informed decisions and sends over the policies back to the \textit{Deceiving Hooks}. \circled{8}~Consequently, the target app receives the deceived data rather than the original data from the hook method. \circled{9}~Hence, the target app sends deceived data over to its server, which ensures user data privacy. 

Additionally, the installed hooks also log the various actions performed by the hooked application with respect to user data access. \circled{10}~These logs are then relayed back to \framework{} when \textit{Deceiving Hooks} communicates with the \textit{Deceiving Module} during hooking. \circled{11}~These logs are shared with the user by the \textit{Resource Access Log Reporter} for manual investigation by the user. Figure \ref{fig:frmwrk-ui}b presents a screenshot of the \textit{Resource Access Log Reporter} logs recorded during the experiments.

%% file: 6_mitigating_side_channel_attacks.tex
\section{Defending against Malware Attacks}
\label{sec:mitigating_sca}


In this section, we show that \framework offers a novel approach to mitigate the
risk of leaking sensitive user data.

\subsection{Testing \framework{} against Real-world Malicious Apps}
\label{sec:malicious_apps}

To analyze the defense capabilities of \framework
against real-world malicious apps, we selected 30 most downloaded apps 
that had been recently banned from Google Play
Store. Due to the absence of accessible downloads for banned apps on the Google Play
Store, we downloaded the latest versions available on third-party platforms. 
We then scrutinized the malicious behavior of these apps using the sensitive API
access logs captured by the \framework.

\input{tables/malicious_apps}

Table \ref{tab:malicious_apps} summaries the list of malicious behaviors we found 
in the 30 malicious apps. We expect users to be able to find malicious behaviors
in a similar manner. In future, we plan to extend \framework to notify users upon 
suspected malicious behaviors.

Malicious behaviors fell under 6 categories:

\noindent\textbf{\textit{Upload sensitive data to the internet.}} Some apps were
identified for transmitting substantial amounts of data over the internet while
operating in the background. For example, we suspect \textit{PhoneFinder by
Clapping} was uploading user audio data as it had also been granted audio
permission. We could not revoke or spoof audio permission without hurting the
app's functionality. Consequently, \framework limited the app's internet access
to safeguard user privacy.

\noindent\textbf{\textit{Accessing sensitive data unnecessarily.}} A typical
pattern among malicious apps was to try to access irrelevant sensitive data. For
example, \textit{Amazing Video Editor} and \textit{Keyboard Themes} apps were
accessing user's contacts information. Similarly, \textit{Instant Speech
Translation} was accessing user's photos. Denying requests for these permissions
led to crashes or termination of the apps. \framework could successfully spoof
such unnecessary sensitive data accessed by these apps without limiting app
functionality.

\noindent\textbf{\textit{Accessing data without user knowledge.}} Other apps 
were calling sensitive APIs at unexpected times. For example, \textit{Free
Translator Photo} app translates images uploaded by the user. However, the app
was found to attempt access to user images even when user had not requested a
translation. \textit{Bus Driver Simulator} allows users to play a game using
accelerometer and gyroscope sensor data. But the app was accessing the sensors
while it was in background. \framework could effectively deceive all the
sensitive data including camera, audio, contacts, location, sensors, and device
information. Users can easily configure deceiving policies to deceive only while
the app is in background while allowing sensitive data while the app is in
foreground.

\textit{GeoSpot: GPS tracker} can let users share their current location with
friends. However, it was found to track the user even when the user has not
asked the app to share their location. \framework could effectively deceive the
location of the user. But since both regular and malicious operations happen
while app is in background, user has to enable/disable deceiving manually with
\framework.



\noindent\textbf{\textit{Sending SMS messages.}} Some apps like \textit{Private
SMS} call \texttt{SMSManager.sendTextMessage()} without user consent.
Multiple media reports suggest that such apps commit financial fraud by
subscribing users to premium services through SMS activation. \framework
prevented sending these messages from the device. However, doing so also limits
the app's functionality. \framework is therefore not effective where the app
does malicious activities using the data needed for its core functionalities.

\subsection{Mitigating Side-Channel Attacks}
\label{sec:side_channel_attack}
Side-channel attacks present a significant threat to the security of mobile devices. These attacks exploit unintentional channels of information leakage, such as accelerometers, gyroscopes, power consumption, or electromagnetic radiation, to infer sensitive data. With their abundance of sensors and communication interfaces, mobile devices are particularly vulnerable to side-channel attacks. These attacks can lead to the unauthorized disclosure of valuable information like passwords, cryptographic keys, or biometric data.

Recent studies~\cite{hasan2013sensing, simon2013pin, ba2020learning, shen2015input} have shed light on the risks associated with so-called \textit{Normal} Permissions, which include seemingly innocuous actions like reading sensor data that do not require explicit user approval. However, in various attacks on user privacy, these permissions are leveraged. 

In response to the pressing privacy concern, our framework offers a novel approach to mitigate the risk of side-channel attacks perpetrated by malicious apps. By selectively modifying permissions granted to specific apps, \framework{} fortifies user privacy and bolsters mobile device security. \framework{} achieves this by deceiving user data, effectively disrupting the ability of malicious apps to gather sensitive information through side-channel channels.

We showcase the effectiveness of the \framework{} in countering the insidious side-channel attacks, by conducting tests employing \textit{GyroSec}~\cite{lin2019motion}. \textit{GyroSec} is an Android application that captures readings from the device's accelerometer and gyroscope without the user's knowledge or consent. This unauthorized data is then transmitted to a remote server where \textit{GyroSec} employs machine-learning algorithms to discern and predict the precise location of touch inputs on the screen. This capability poses a significant danger, as it leads to leakage of users' sensitive information. Our experiments have uncovered that \textit{GyroSec} can carry out such side-channel attacks without any restrictions from the Android Permission Framework.

\begin{figure}[t]
\centering
\begin{subfigure}{0.4\linewidth}
    \includegraphics[width=\linewidth]{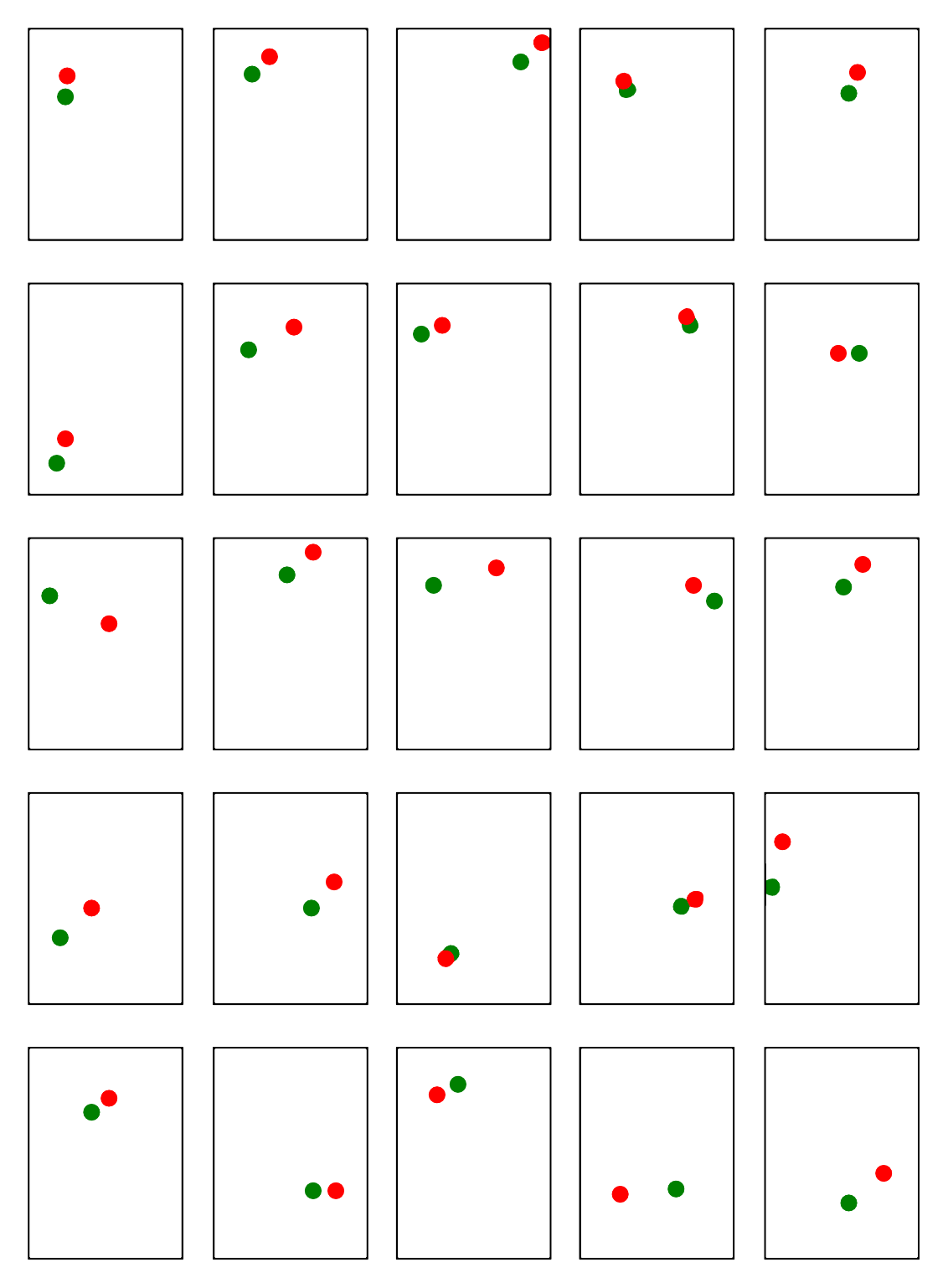}
    \caption{Without \framework{}}
    \label{fig:tchPredict_wo_frmwrk}
\end{subfigure}
\hspace{0.9mm}
\begin{subfigure}{0.4\linewidth}
    \includegraphics[width=\linewidth]{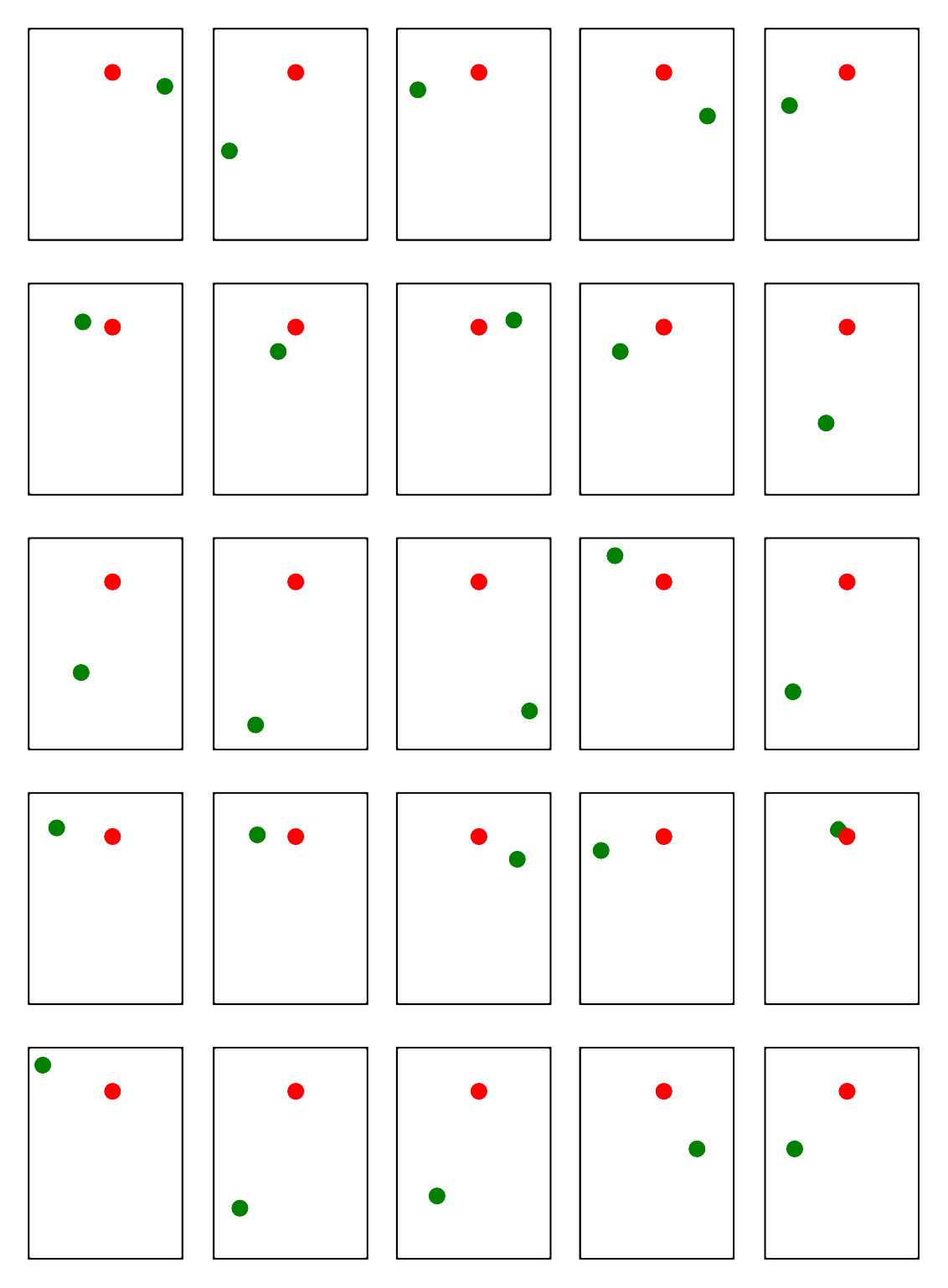}
    \caption{With \framework{}}
    \label{fig:tchPredict_w_frmwrk}
\end{subfigure}
\caption{Touch predicitions made by \textit{Gyrosec} server based on sensor readings received.}
\label{fig:tchPredict}
\end{figure}

By configuring \framework{} to deceive \textit{GyroSec}, we logged every instance of data access by \textit{GyroSec}, promptly notifying users of any background resource data breaches. \framework{} further deceived the data received by \textit{GyroSec}. Using the \textit{Policy Configurator} the accelerometer and gyroscope sensor readings were deceived as constant value for Gyrosec.
Figure \ref{fig:tchPredict} illustrates reduction in \textit{GyroSec}'s ability to predict accurately by \framework{}. The prediction accuracy dropped from \textbf{81.22\%} to \textbf{5.36\%}, proving \framework{} to be a successful measure against Permission-based Side-Channel Attacks.

\textit{Overall, this section shows that \framework could effectively stop most 
malicious behaviors giving better control to users over their sensitive data.}


%% file: tables/malicious_apps.tex
\begin{table}
    {\centering
    \begin{tabular}{p{\linewidth}} 
         \toprule
         \textbf{Malicious behavior:} Upload sensitive data to internet\\
         \textbf{Apps (downloads):}
         All Good PDF Scanner (10M+), Fast PDF Scanner (5M+), PhoneFinder by
         Clapping (5M+), What's Me Sticker (1M+)\\
         \textbf{\framework:} Limit internet access to the app\\
         \midrule
         \midrule
         \textbf{Malicious behavior:} Access sensitive user data like Contacts, Location, Audio,
         and Sensors unnecessarily\\
         \textbf{Apps (downloads):} Amazing Video Editor (5M+), CapCut Pro (5M+), Instant Speech Translation (5M+), Keyboard Themes (5M+),
         Launcher iOS 15 (5M+) \\
         \textbf{\framework:} Deceived all the unnecessary user data requested without crashing the apps\\
         \midrule
         \midrule
         \textbf{Malicious behavior:} Accessing data without user knowledge\\
         \textit{Accessing sensor data in the background}\\
         \textbf{Apps (downloads):} Bus Driver Simulator (5M+), Bus - Metrolis 2021 (1M+),
         Fingerprint Changer (1M+), Fingerprint Defender (1M+), Lifeel - scan and test (5M+),
         Locker Tool (5M+), OFFRoaders - Survive (5M+), Racers Car Driver (5M+), Safe Lock (5M+),
         Slime Simulator (5M+), Smart Spot Locator (1M+), Unique Keyboard (5M+) \\
         \textbf{\framework:} Allowed the apps to access the sensor
         data in the foreground but not in the background\\
         \midrule
         
         \textit{Camera access without user consent and knowledge}\\
         \textbf{Apps (downloads):} Free Translator Photo (5M+), Handy Translator Pro (10M+),
         Heart Rate and Pulse Tracker (5M+), Heart Rhythm (1M+),
         My Chat Translator (5M+) \\
         \textbf{\framework:} Deceived camera data fed into the app\\
         \midrule
         
         \textit{Location continuously tracked by app}\\
         \textbf{Apps (downloads):} Geospot: GPS Tracker (5M+), iCare - Find Location (5M+) \\
         \textbf{\framework:} Deceived location data according to user policy\\
         \midrule
        \midrule
        \textbf{Malicious behavior:} Detected accessing Send SMS API without user knowledge\\
        \textbf{Apps (downloads):} Private SMS (5M+), Mint Left Messages (1M+) \\
        \textbf{\framework:} Blocked access to send SMS but not from reading SMS as per user policy\\
        \bottomrule
    \end{tabular}
    }
    \caption{Malicious behavior of Android apps banned by Google and the actions taken by \framework 
    to protect user privacy.}
    \label{tab:malicious_apps}
\end{table}

%% file: 5_protecting_user_privacy.tex
\section{Privacy Protection against Real-World Apps}
\label{sec:protecting_up}

In Figure~\ref{fig:intro-case-study-uber}, we showed how \framework can be
effectively used to protect user privacy while they are using Uber.  In this
section, we use three more real-world Android apps to showcase privacy
advantages of \framework.

\subsection{com.katana.facebook (v404.0.0.35.70)}
\label{sec:fb_case_study}

\begin{figure}[t]
    \centering
    \includegraphics[width=\linewidth]{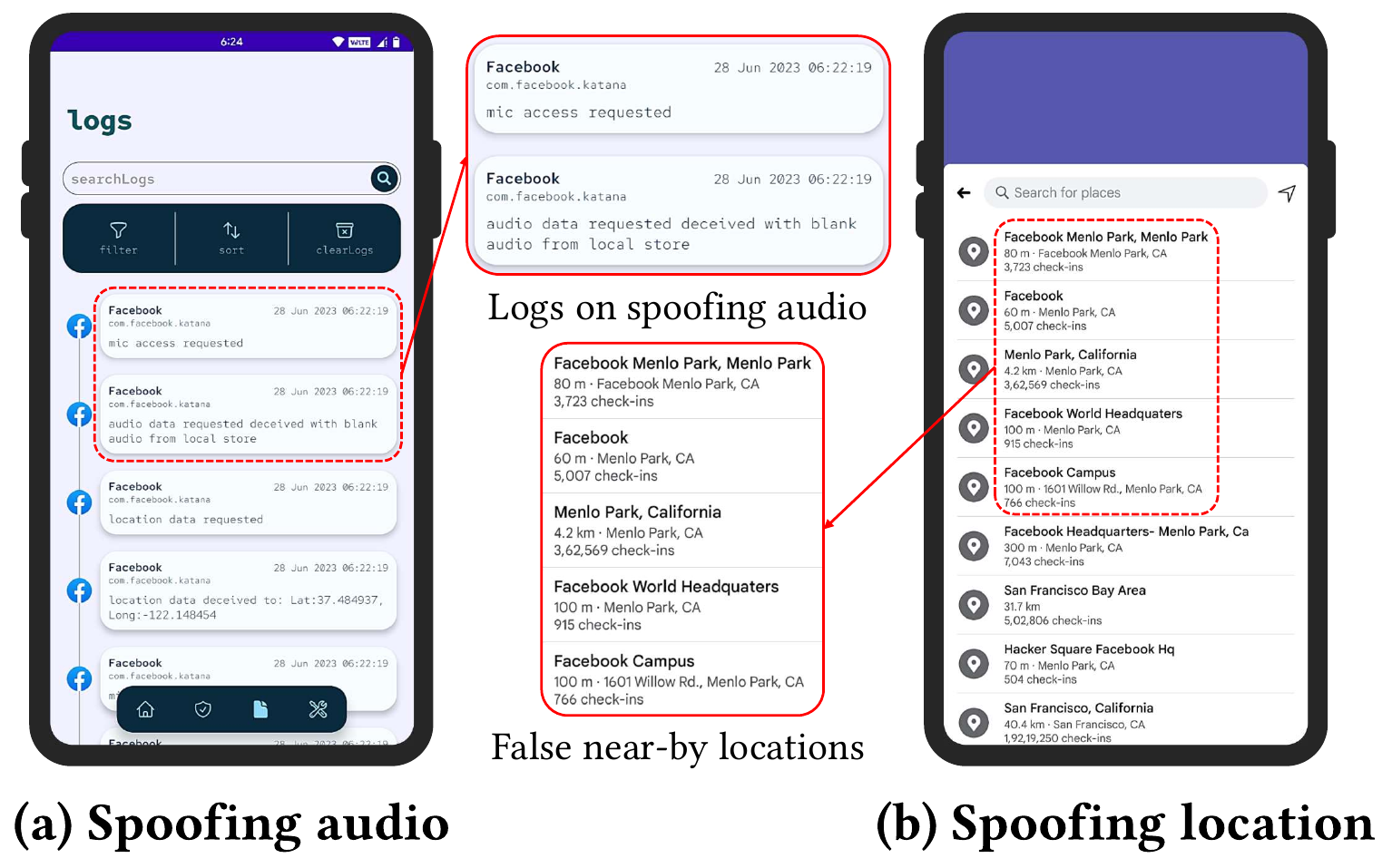}
    \caption{Screenshots demonstrating deceiving location and audio data using \framework{} for the Facebook app.}
    \label{fig:case-study-facebook}
\end{figure}

During the experiment, while scrolling through posts in the Facebook app's
\textit{Home} activity and without engaging any other app features,
\framework's \textit{Resource Access Log Reporter} recorded instances of the
Facebook app accessing audio permission as shown in
Figure~\ref{fig:case-study-facebook}(a). On further investigation, we found
that the Facebook app was using \texttt{MediaRecorder} to access the device
microphone. Moreover, the Android's privacy indicator~\cite{andPrivacyIndicator}
(green dot), which indicates active use of the device microphone, did not appear
when \framework{} logged the use of audio permission. We found this surprising.

We were able to repeat this behavior with a custom mobile app \ie we used
\texttt{MediaRecorder} API to successfully record audio
while Android's privacy indicator did not show up. We believe it is a bug in
Android's privacy indicator implementation. We have reported this to Android and
plan to open source this custom app responsibly.

Hence, \framework{} was able to detect microphone usage that was not indicated
by Android. Moreover, \framework{} successfully spoofed audio data requested
by the Facebook app. 

Similar to the unexpected microphone access, \framework{}
also logged instances of calendar data access by the Facebook app while
searching for nearby events. The app reads user's calendar events to display their
availability when the user is viewing nearby events.  
Users can configure \framework to manipulate calendar data in a way that allows
the Facebook app to receive information about their availability without
compromising sensitive details about individual events. This involves spoofing
fields like event name and location. 
\framework could spoof the app's request for calendar data by providing a list
of manipulated events.

Nearby events were listed using user's location
data. As shown in Figure~\ref{fig:case-study-facebook}(b), we could
effectively spoof the fine-grained GPS location data with a nearby location
ensuring that the reported nearby places and events were same without
compromising user's exact location. 

\subsection{com.snapchat.android (v12.24.0.34)}
\label{sec:sc_case_study}

\begin{figure}[t]
    \centering
    \includegraphics[width=\linewidth]{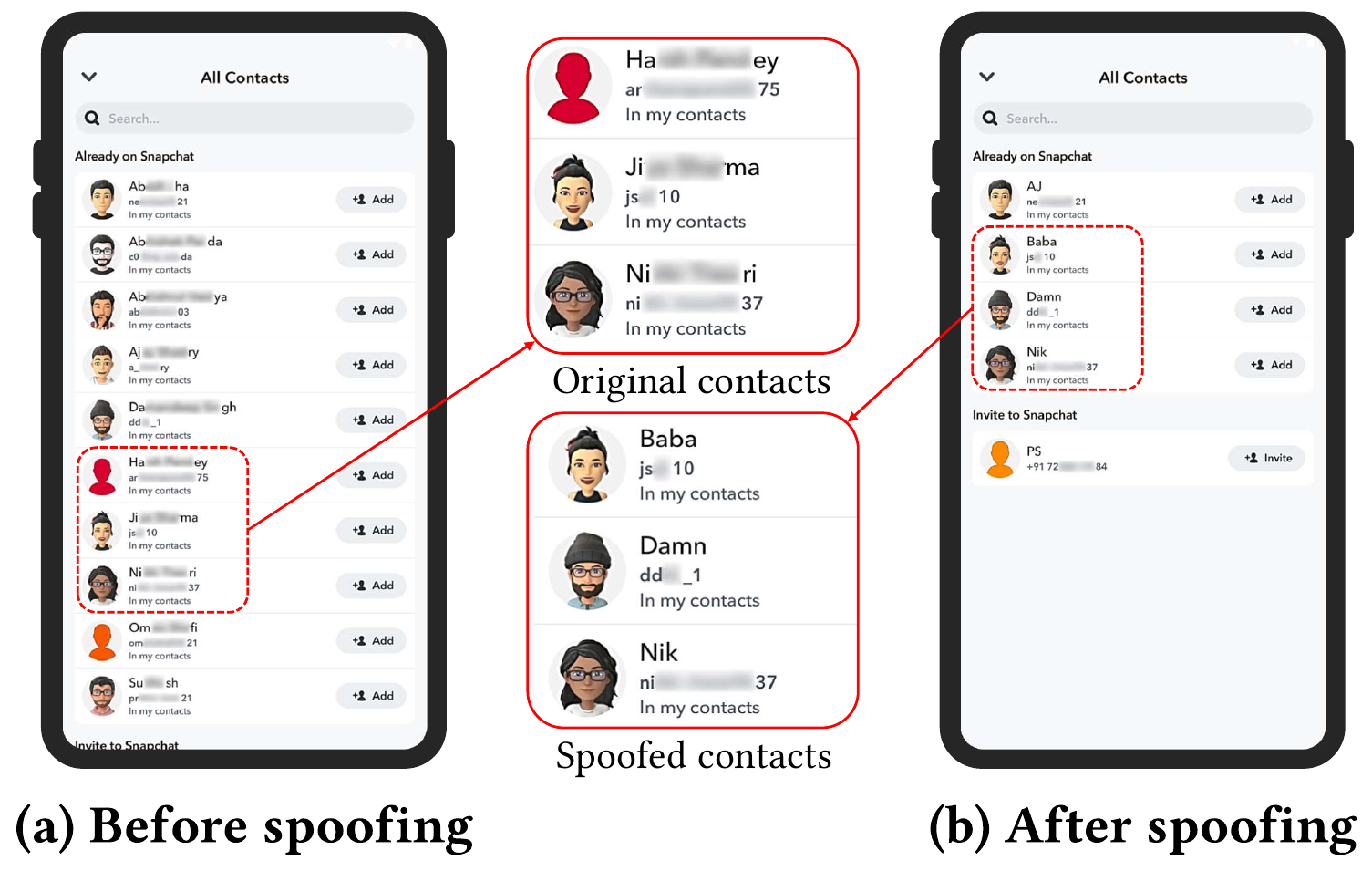}
    \caption{Screenshots demonstrating how user can share spoofed contacts using WhiteLie with Snapchat.}
    \label{fig:case-study-snapchat}
\end{figure}


Snapchat has a feature called 
\textit{Snap Map}, which utilizes the device's GPS and other sensors to
display user avatars on a map, showcasing real-time physical activities. 
While Snapchat does provide \textit{Ghost Mode} as
an option to refrain from sharing location and sensor data, it sacrifices
\textit{SnapMap} features that rely on this information. Furthermore,
\textit{Ghost Mode} does not guarantee that the app is not accessing the data. 


\framework successfully spoofed device's GPS location and
\texttt{READ\_CONTACTS} by utilizing a list of contacts specified in
\framework's \textit{Deceit} section as shown in Figure
\ref{fig:case-study-snapchat}(b). Users can thus limit the information 
they want to share with Snapchat and still use Snap Map feature.

\subsection{com.truecaller (v12.54.7)}
\label{sec:tc_case_study}
\begin{figure}[t]\centering
    \includegraphics[width=\linewidth]{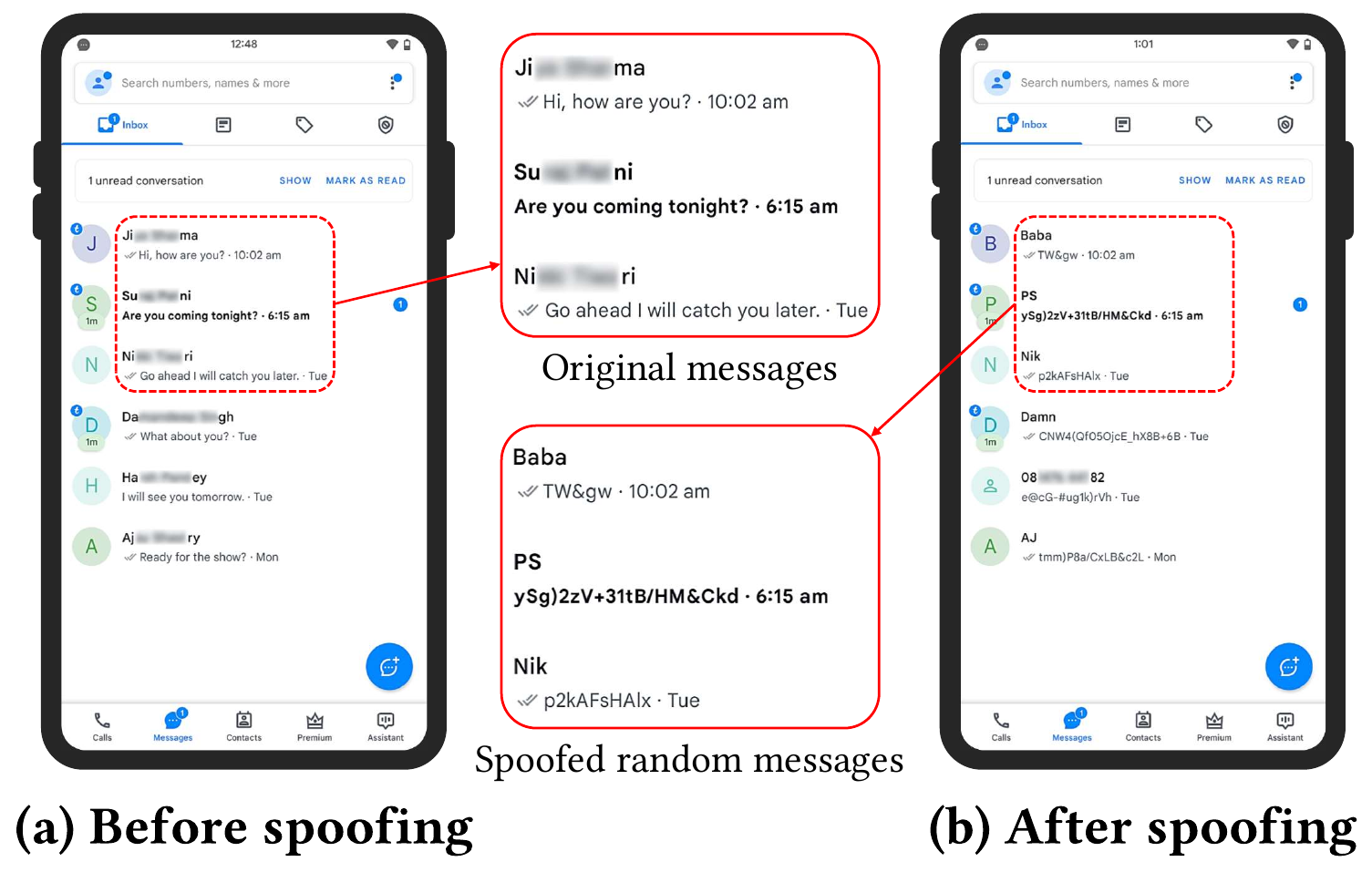}
    \caption{Screenshots demonstrating how a user can share spoofed SMS using WhiteLie with Truecaller app.}
    \label{fig:case-study-truecaller}
\end{figure}

Truecaller provides sender information about incoming text messages to highlight
known spammers. With Truecaller, users can ignore spam messages. But to use
this feature, users have to give permission to the app to read all their
messages. 
We used \framework to successfully spoof message content while retaining the phone 
number from which the message was received.
This way users can retain control over the privacy of their messages 
while still identifying whether senders are known spammers as shown in
Figure~\ref{fig:case-study-truecaller}(b).

\textit{Overall, this section shows that \framework empowers users to fully use
the app features while controlling the information that they share with the
apps.}

%% file: 7_results.tex
\section{Evaluating Overhead}
\label{sec:results}

In this section, we present the outcomes of our evaluation of \framework{},
focusing on its performance overhead in a real device while running various real
world apps.  We first quantify per API call overhead of \framework using
micro-benchmarks. And then evaluate memory and battery overhead of \framework{}
was evaluated using 10 most popular apps downloaded from the Google
Play Store.

All permissions requested by the apps were granted beforehand, and the apps were
subjected to 10 minutes of manual usage 5 times. All experiments were conducted
on a \textit{Samsung Galaxy M21} smartphone with a 2.3 GHz octa-core processor
and 4 GB of RAM running Android 12.

\subsubsection*{\textbf{API call overhead}} 
We measure the time elapsed during code execution by executing hooked
API methods within the \texttt{Timing.measureNanoTime()} method.
We observed an average overhead of 1.64 ms across different API calls.
The increase in elapsed time is because of the operations performed by the
hooks. For user data like contacts and camera, we experienced an overhead of
3.87 ms and 3.23 ms respectively. For tracking and clipboard user data, we
observed an overhead of 0.83 ms and 1.07 ms respectively. The overhead depends
on the operations performed by the respective permission-deceiving hooks. For
contacts and camera APIs, we are performing intensive operations like reading a
database and manipulating pixels of an image, whereas for tracking and clipboard
APIs, we are simply returning a deceived string.

\begin{figure}[t]
\includegraphics[width=0.8\linewidth]{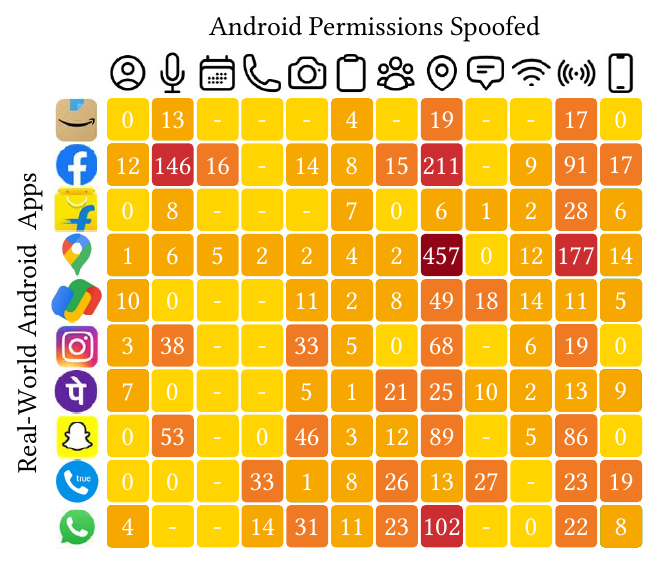}
\caption{Heatmap illustrating number of API calls made by various popular apps while evaluating performance overhead.}
\label{fig:reslts_noAPICalls}
\end{figure}

Many of these APIs were called several times during the app runs in our
experiments as shown in Figure~\ref{fig:reslts_noAPICalls}. The API call
overhead, however, did not impact the performance of the apps we tested in any
observable manner.

\subsubsection*{\textbf{Memory Used}} 
The memory utilization of the benchmarking app was measured using \textit{ADB}'s
\textit{dumpsys} tool, using the \textit{Proportional Set Size (PSS)} metric.
This metric captures the shared memory proportionally used by each
process.


\begin{figure}[t]
\includegraphics[width=\linewidth]{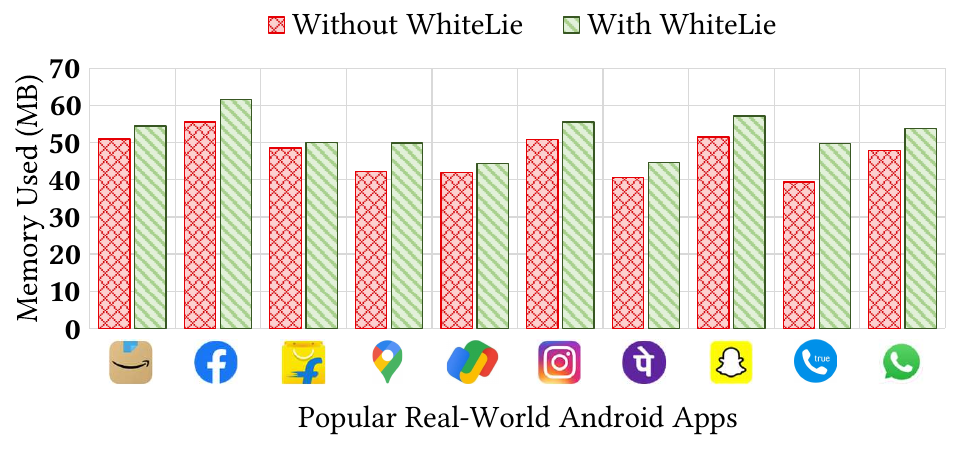}
\caption{Memory used by various Android apps with and without \framework{}.}
\label{fig:results_memUsedAll}
\end{figure}




Figure~\ref{fig:results_memUsedAll} shows that the memory overhead while using
\framework{} on various real-world apps is negligible.  On average, \framework
only incurs 5.2 MB of memory overhead. 


\subsubsection*{\textbf{Battery Discharged}} 
To accurately measure the device's battery usage during our experiments, we
utilized the \textit{ADB}'s \textit{batterystats} tool~\cite{batterystats}.
Before each experiment, we ensured a clean slate by resetting the tool. To
mitigate any potential inconsistencies in battery usage statistics, we remotely
connected the device via \textit{Wifi ADB}, as physically connecting it to a PC
can inadvertently charge the device and yield inaccurate results. 
By leveraging \texttt{batterystats}'s \textit{discharge} metric, we could
quantify the amount of battery discharged since the last charge, accounting for
the impact by both the target app and the system. 

\begin{figure}[t]
\includegraphics[width=\linewidth]{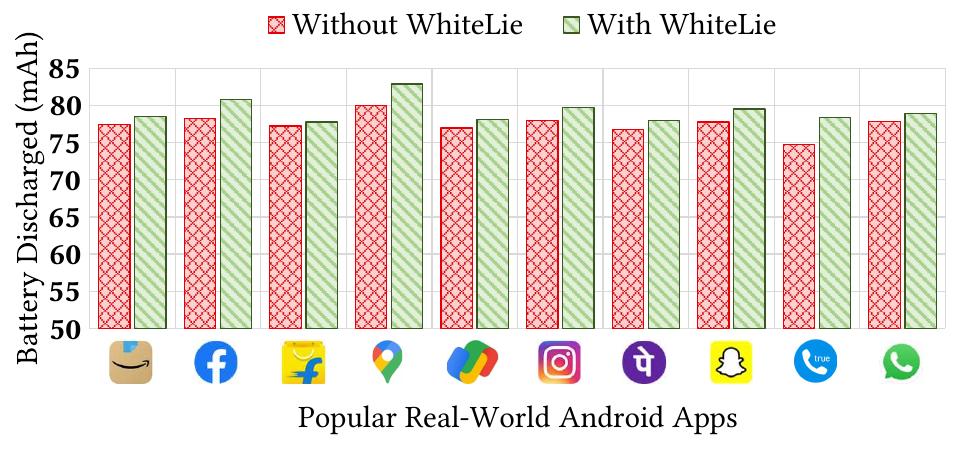}
\caption{Battery discharged by various Android apps with and without \framework{}.}
\label{fig:reslts_btryDschrgd}
\end{figure}

Figure \ref{fig:reslts_btryDschrgd} shows the battery drained during the 
app runs with and without \framework. It shows that \framework's impact on
battery drain is negligible. An average additional discharge of approximately
1.76 mAh (2.52\%) was observed across the 5 minute runs of various apps experimented when
they were run with \framework. 

\textit{Overall, our experiments show that \framework's usage has a minimal impact on
app's performance, app's memory consumption, and the device's battery drain. 
This shows that \framework is a versatile and efficient approach to protecting
user privacy.}

We further experimented with actively reducing battery drain using \framework.
In particular, by returning \texttt{null} from the
\texttt{beforeHookedMethod()}, the method calls to the original method can be
blocked which can save battery consumed by sensors and other components. To 
examine this, we added a battery saver mode to \framework{} that selectively
blocks Android API calls according to user policies. 
To quantitatively measure the battery drainage, we developed a benchmarking app that performed a consecutive series of 1000 API calls to retrieve user data.
We observed this approach can save an average of 4.08 $\mu{}Ah$
(60.83\%) per API call.

\begin{figure}[t]
\includegraphics[width=\linewidth]{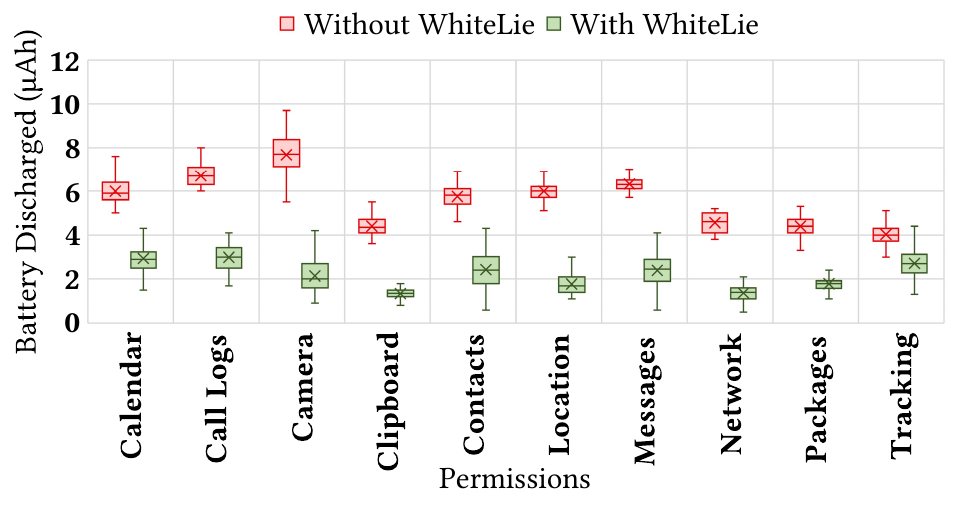}
\caption{Battery discharged by benchmarking app while fetching user data for various Permissions with and without \framework{}.}
\label{fig:reslts_btrySaver}
\end{figure}



%% file: 9_conclusion.tex
\section{Conclusion}
\label{sec:conclusion}

In conclusion, this paper introduces \framework{}, a robust user data spoofing system designed to enhance user privacy without compromising app functionality on non-rooted Android devices with minimal overhead. Through extensive evaluations on popular apps \framework{} demonstrated its ability to effectively spoof 78.32\% of requested permissions without detection or app crashes. Beyond its primary purpose, \framework{} revealed its potential to bypass continuous authentication mechanisms based on sensor data, highlighting vulnerabilities in existing security measures. The findings of this research emphasize the importance of improving ongoing authentication methods and enhancing privacy safeguards and security measures within the Android ecosystem. The system also proved effective in detecting unexpected permission accesses by benign apps and mitigating side-channel attacks launched by malicious apps. Real-world examples, such as Facebook's unauthorized microphone access and protection against recently banned apps like Private SMS, showcase the diverse applications of \framework{}. 
